\documentclass[12pt,preprint]{aastex}
\usepackage{epsfig}
\usepackage{lscape}

\makeatletter

\newcommand{\Rmnum}[1]{\expandafter\@slowromancap\romannumeral #1@}
\makeatother

\def\Lsun{\hbox{\it L$_\odot$}}

\def\Msun{\hbox{\it M$_\odot$}}
\def\Zsun{\hbox{\it Z$_\odot$}}

\newcommand{\etal}{\mbox{et al.}}

\newcommand{\program}[1]{{\tt {#1}}}
\def\program{\texttt}
\def\ca{\citeauthor}
\def\cy{\citeyear}

\begin{document}
\title{Isolated Wolf-Rayet Stars and O Supergiants in the Galactic Center Region Identified via Paschen-$\alpha$ Excess}

\shorttitle{}

\author{J.~ C. Mauerhan\altaffilmark{1}, A. Cotera\altaffilmark{2}, H. Dong\altaffilmark{3}, M.~R. Morris\altaffilmark{4}, Q.~D. Wang\altaffilmark{3},  S.~R. Stolovy\altaffilmark{1}, C. Lang\altaffilmark{5}}
\altaffiltext{1}{Infrared Processing and Analysis Center, California Institute of Technology, Mail Code 220-6, 1200 East California Boulevard, Pasadena, CA 91125, USA; mauerhan@ipac.caltech.edu} 
\altaffiltext{2}{SETI Institute, 515 North Whisman Road, Mountain View, CA 94043, USA}
\altaffiltext{3}{Department of Astronomy, University of Massachusetts, Amherst, MA 01003, USA} 
\altaffiltext{4}{Department of Physics and Astronomy, University of California, Los Angeles, CA 90095, USA}
\altaffiltext{5}{Department of Physics and Astronomy, University of Iowa, Iowa City, IA 52245, USA}

\begin{abstract}
We report the discovery of 19 hot, evolved, massive stars near the Galactic center region (GCR). These objects were selected for spectroscopy owing to their detection as strong sources of Paschen-$\alpha$ emission-line excess, following a narrowband imaging survey of the central $0\fdg65\times0\fdg25$ $(l, b)$ around Sgr A$^*$ with the \textit{Hubble Space Telescope}.  Discoveries include five carbon-type (WC) and six nitrogen-type (WN) Wolf-Rayet stars, six O supergiants, and two B supergiants.  Two of the O supergiants have X-ray counterparts, the properties of which are broadly consistent with colliding-wind binaries and solitary O stars. The infrared photometry of 17 stars is consistent with the Galactic center distance, but two of them are located in the foreground. Several WC stars exhibit a relatively large infrared excesses, which is possibly the signature of thermal emission from hot dust. Most of the stars appear scattered throughout the GCR, with no relation to the three known massive young clusters; several others lie near the Arches and Quintuplet clusters and may have originated within one of these systems.  The results of this work bring the total sample of Wolf-Rayet stars in the GCR to 92. All sources of strong P$\alpha$ excess have been identified in the area surveyed with \textit{HST}, which implies that the sample of WN stars in this region is near completion, and is dominated by late (WNL) subtypes. The current WC sample, although probably not complete, is almost exclusively dominated by late (WCL) subtypes. The observed Wolf-Rayet subtype distribution in the GCR is a reflection of the intrinsic rarity of early subtypes (WNE and WCE) in the inner Galaxy, an effect that is driven by metallicity.
\end{abstract}

\section{Introduction}
The Galactic center region (GCR) has the highest star formation rate density in the Milky Way, more than $\approx$2 orders of magnitude above the average value of the Galactic disk (\citeauthor{fig04}~\citeyear{fig04}). This is considered to be a consequence of the large reservoir of molecular gas that fills the central half-kiloparsec of the Galaxy, where $\approx$10\% of the Galaxy's molecular gas is concentrated. With respect to the physical properties that are typical of molecular gas further out in the Galactic disk, the GCR medium is significantly hotter, more turbulent, and more highly magnetized. It is also subject to mechanical shearing by the strong tidal field, and repeated compression from cloud-cloud collisions, stellar winds, and supernova shock waves (\citeauthor{mor96}~\citeyear{mor96}). It has been suggested that the environmental conditions in the GCR result in an elevated Jeans mass and an enhanced rate of massive-star production. 

At least three extraordinary clusters of massive stars have formed within the unique GCR medium during the past several Myr, including the Arches cluster (\ca{nag95}~\cy{nag95}) and the Quintuplet cluster (\ca{oku90}~\cy{oku90}; \ca{nag90}~\cy{nag90}; \ca{glass90}~\cy{glass90}), both of which lie $\approx$30 pc, in projection, from the center. Within the inner few parsecs of the Galaxy lies the Central cluster (\ca{krabbe91} ~\cy{krabbe91}), which is gravitationally bound to the Galactic black hole, Sgr A$^*$ (\ca{scho02}~\cy{scho02}; \ca{ghez03}~\cy{ghez03}). All three of the known clusters are massive ($M\sim10^4 \Msun$), dense, containing hundreds of O stars, and dozens of evolved Wolf-Rayet stars (WRs) and OB supergiants (\ca{fig99a}~\cy{fig99a}, \cy{fig02}; \ca{pau06} \cy{pau06}). The collection of three such extraordinary clusters in the GCR is evidence favoring the hypothesis that the predominant mode of star formation in the region gives rise to dense, massive clusters. However, surveys aimed at identifying massive stars throughout the greater GCR have revealed a population of several dozen massive stars that have no apparent spatial association with the known stellar clusters (\ca{cot99}~\cy{cot99}; \ca{hom03}~\cy{hom03}; \ca{mun06}~\cy{mun06}; \ca{mik06}~\cy{mik06}; \ca{mau07}~\cy{mau07}; \ca{mau10a}~\cy{mau10a}, \cy{mau10b}). The discovery of  these ``isolated" massive stars has motivated the continued search for additional such cases, with the goal of determining the total extent of the inter-cluster population of massive stars and their physical relation to, or independence from, starburst clusters like the Arches and Quintuplet.  The demography and kinematics of these isolated stars will yield valuable insight into the mode of massive star formation in the GCR and the dynamical evolution of massive stellar clusters within galactic nuclei.  

The identification of isolated massive stars in the GCR is a difficult task. Heavy extinction limits observations of the Galactic center starlight to the infrared part of the electromagnetic spectrum. Infrared measurements sample the Rayleigh-Jeans tail of stellar spectral energy distributions where there is a degeneracy of stellar color with effective temperature. This renders early-type and late-type stars photometrically indistinguishable.  Narrowband photometry aimed at detecting emission-line excesses from hot supergiants can be very effective but requires a highly stable point-spread function (PSF) in order to accurately subtract the stellar continuum emission from the spectral line emission.  This is difficult to achieve in seeing limited conditions, especially when extreme stellar confusion is an issue, as it typically is when observing the GCR. Therefore, infrared narrowband surveys are best conducted from the stable environment of space, where low-background, diffraction-limited observations can be executed without the nuisance of PSF variability. Furthermore, space-based narrowband observations enable measurement of the Paschen-$\alpha$ (P$\alpha$) transition ($n$=4--3) of H {\sc i}, which is the strongest emission line of this species in the infrared. The $\lambda$1.87 {\micron} wavelength of this transition is also shared with several transitions of He {\sc i} and He {\sc ii}, which is important in the interest of detecting hydrogen-poor WRs that exhibit strong helium emission.  

In this work, we report discoveries of ``isolated" emission-line stars in the GCR via P$\alpha$ narrowband imaging from space and subsequent infrared spectroscopy from the ground.  The new additions add to our recent identification of a new luminous blue variable in the GCR (\ca{mau10b}~\cy{mau10b}), which was the first reported discovery of the spectroscopic survey. The successful detection of these stars highlights the significant advantage that space-based narrowband searches using the P$\alpha$ line have over ground-based surveys centered on other atomic transitions (e.~g., \ca{fig95}~\cy{fig95}; \ca{hom03}~\cy{hom03}) which have identified only a fraction of the isolated emission-line stars.  We present here the spectra of new identifications, their photometric properties, and discuss the implications that these objects have for massive star formation in the GCR. 

\section{Observations}
\subsection{Selection of P$\alpha$ Emission-line Stars}
Our target sample was assembled following the completion of a P$\alpha$ narrowband imaging survey of the GCR, performed with the \textit{Hubble Space Telescope} and the Near-Infrared Camera and Multi-Object Spectrometer (NICMOS; \citeauthor{wang10}\citeyear{wang10}). The survey, requiring 144 orbits of observations, covered the central $(l, b)\approx39\times15$ arcmin$^{2}$ around Sgr A$^*$, which corresponds to a total projected physical area of $\approx90\times35$ pc$^2$, assuming a distance of 8 kpc to Sgr A$^*$ (\ca{reid93}~\cy{reid93} and references therein). The NIC3 camera was utilized and narrowband images were obtained through the F187N (line; $\lambda=1.874$ {\micron}; FWHM=0.019 {\micron}) and F190N (continuum; $\lambda=1.901$ {\micron}; FWHM=0.017 {\micron}) filters. The survey resulted in the detection of $\approx$150 point-like sources of P$\alpha$ line excess (i.e., sources with F187N-F190N values $\gtrsim$5$\sigma$ above the mean of all point sources in the survey area; Dong et al., in preparation). Approximately half of the sources are located outside of the three known stellar clusters, and nearly half of those were unidentified prior to this work. 

We have undertaken a near-infrared spectroscopic campaign to characterize the unidentified sources of  P$\alpha$ line excess. Focusing on the sources having the strongest line excess and the  brightest near-infrared counterparts ($K_s\lesssim12.7$ mag), we have spectroscopically identified 19 new, hot, massive stars in the GCR.  So far, our spectroscopic search has had a 100\% success rate for detecting emission-line stars. Table 1 lists basic data for these sources, including their positions, corresponding $JHK_s$ photometry, and narrowband F187N and F190N measurements. The sources are named based on their Galactic coordinates, but will hereafter be referred to by their numerical entry in Table 1. The positions and $JHK_s$ photometry were taken from the catalog of the SIRIUS survey of the GCR (see \ca{nish06}~\cy{nish06} and references therein), which generated a catalog of point sources having a 10$\sigma$ limiting $K_s$ magnitude of 15.6.  The F187N and F190N measurements are from Dong et al. (in preparation). We note that the method of extraction for the narrowband photometry has been refined since our report on the LBV G0.120$-$0.048 (\ca{mau10b}~\cy{mau10b}), where we presented our narrowband measurements for this source and the known LBVs qF362 (aka FMM 362; \ca{geb00}~\cy{geb00}) and the Pistol Star (\ca{fig98}~\cy{fig98}); so, we take this opportunity to include the refined narrowband measurements for all three of these LBVs in Table 1.

\subsection{Spectroscopic Observations}
Spectra of four stars were obtained using the SpeX medium-resolution spectrograph on the 3 m Infrared Telescope Facility (IRTF) telescope (\ca{ray03}~\cy{ray03}), located on the summit of Mauna Kea in Hawaii.  SpeX was used in short cross-dispersed mode (SXD), utilizing a slit width of 0\farcs5, providing a spectral  resolution of $R=\lambda/\delta\lambda\approx1200$. Flat-field images and a wavelength calibration source were provided by obtaining spectra of continuum and argon lamp sources. The data were reduced and extracted using the \program{IDL}-based software package \program{Spextool}, specially designed for the reduction of data obtained with SpeX on the IRTF (\ca{cus04}~\cy{cus04}). 

The 3 m United Kingdom Infrared Telescope (UKIRT) on Mauna Kea was used to obtain spectra of six stars. The UKIRT 1--5 {\micron} Imager Spectrometer (UIST; \ca{ram04}~\cy{ram04}) was used in service mode as part of the UKIRT Service Programme. The short-$K$ grism and 4-pixel slit (0\farcs48) were used, providing a spectral resolution of $R\approx2133$ and a wavelength range of $\lambda$2.01--2.26 $\micron$. Flat-fields and a wavelength calibration source were provided by obtaining spectra of continuum and argon lamps. The data images were reduced using the \program{Starlink ORACDR} pipeline. The spectra were extracted using the \program{IRAF} routine \program{APALL}.

Spectra of three stars were obtained using the 4.1 m Anglo-Australian Telescope (AAT) on Siding Spring Mountain (Mount Woorat) in New South Wales, Australia. The IRIS2 instrument (\ca{tin04}~\cy{tin04}) provided a spectral resolution of $R\approx2400$ in the $K$ band, using the 1\arcsec~slit. Flat-field images and a wavelength calibration source were provided by obtaining spectra of continuum and neon lamp sources. The data images were reduced and spectra extracted with the same programs used for the UKIRT data.

Spectra of six stars were obtained at the Southern Observatory for Astrophysical Research (SOAR), located on Cerro Pachon in Chile. The Ohio State Infrared Imager Spectrometer (OSIRIS; \ca{dep93}~\cy{dep93}) was used in cross-dispersed mode for five of these stars. This mode provides a spectral resolution of $R\approx1200$. Wavelength calibration was achieved using the background OH emission lines from the sky. For one star, OSIRIS was used in the higher resolution longslit mode to obtain a $K$-band spectrum with a resolution $R\approx3000$.  For this observation, wavelength calibration was performed using spectra of the internal neon and argon lamps. All of these data were reduced and extracted using the standard \program{IRAF} routines and \program{APALL}.

For all observations the spectra were acquired in an ``ABBA" nodding sequence in order to subtract the sky background, and to suppress the contribution of bad pixels. Spectra of A0V standard stars were obtained at airmasses similar to those of the science targets in order to derive a telluric absorption spectrum. The telluric corrections were applied using the IDL package \program{xtellcor} (\ca{vac03}~\cy{vac03}), which removes model H {\sc i} absorption lines from the A0{\sc V} standard star before application to the science data. Table 2 summarizes the facilities used for the observations of each star. 

\section{Spectroscopic Classification}
In the following sections, the spectroscopic criteria we use to classify these stars were adopted from previous spectroscopic studies of massive stars in the near-infrared, conducted by \ca{pmor96}~(\cy{pmor96}), \ca{fig97}(\cy{fig97}), and \ca{mar08}~(\cy{mar08}). The central wavelengths of all spectral lines referred to in the text are adopted from these references. 

\subsection{Early O Supergiants and Weak-lined WNh Stars}
The $K$-band spectra of stars 6 and 7 are presented in Figure 1. The spectra are dominated by a complex of blended emission lines near $\lambda$2.112--2.115 {\micron}, which includes contributions from He {\sc i}, N {\sc iii}, C {\sc iii} and O {\sc iii}.  Blueward of this line complex are emission lines of C {\sc iv} at $\lambda$2.069 and $\lambda$2.078 {\micron}. Br$\gamma$ emission is marginally detected in the spectrum of star 6, although it may be a background nebular feature. The spectra of both stars exhibit He {\sc ii} absorption at $\lambda$2.189 {\micron}, although it is very near the noise level. These features are consistent with the $K$-band spectra of early O supergiants, specifically in the range of  O4--6{\sc I} (e.~g., see \ca{mar08}~\cy{mar08}).
 
The spectra of  stars 16, 15,  12, 9, and 10 are presented in Figure 2. The spectra of these stars are dominated by emission lines of Br$\gamma$ and the $\lambda$2.112--2.115 {\micron} complex of He {\sc i}, N {\sc iii}, C {\sc iii} and O {\sc iii}. Stars 16, 15, and  12  exhibit the $\lambda$2.112--2.115 {\micron} complex in fairly strong emission, while 9 and 10, instead, exhibit a He {\sc i} absorption line accompanied by a relatively weak emission component.  Weak C {\sc iv} emission is also apparent in the spectra of stars 16, 15, and  12. The Br$\gamma$ lines of all of these stars exhibit an asymmetry on their blue side owing to a contribution from He {\sc i} at $\lambda$2.1647 {\micron}. An absorption line of He {\sc i} at $\lambda$2.058 {\micron} is present for four of the stars, while this feature appears in emission for two of them. Weak N {\sc iii} emission appears at $\lambda$2.247 {\micron} in the spectrum of star 16.  Three of the stars also exhibit weak He {\sc ii} absorption at $\lambda$2.189 {\micron}, while this transition is not detected in the spectra stars 9 and 10. These features are characteristics of OIf$^+$\footnote{Historically, the ``+"  in the OIf$^+$ designation is given to O stars exhibiting S {\sc iv} emission ($\lambda\lambda$4089 and 4116 {\AA}), as well as N {\sc iii} and He {\sc ii} $\lambda$4686 {\AA} emission in their optical spectra (\ca{wal71}~\cy{wal71}).} supergiants and late-type WN stars (WNL), which are commonly referred to as hydrogen-rich WRs (WNh stars). Distinguishing between these two types can be very difficult indeed. According to \ca{mar08}~(\cy{mar08}), OIf$^+$ stars exhibit Br$\gamma$ emission that has approximately equivalent strength to the $\lambda$2.112--2.115 {\micron} complex, while WN8--9h stars exhibit a relatively dominant Br$\gamma$ emission line.  Based on the dominance of Br$\gamma$ in the spectra of all stars in Figure 2, we may be inclined to classify them all as WN8--9h stars. However, the weakness of the $\lambda$2.112--2.115 {\micron} complex in the spectra of stars 9 and 10 is more typical of the O4--6If$^{+}$ stars in \ca{mar08}~(\cy{mar08}) than the WN8--9h stars, which always have a strong $\lambda$2.112--2.115 {\micron} emission complex. Furthermore, WN8--9h always exhibit He~{\sc ii} emission or absorption at $\lambda$2.189 {\micron}, but this transition is not detected in the spectra of stars 9 and 10. Thus, we classify stars 16, 15, and  12 as WN8--9h stars, and assign the O4--6If$^+$ designation to stars 9 and 10. 

\subsection{Other OB Supergiants}
The spectra of stars 13 , 18, 5, and 1 are presented in Figure 3. All of these stars exhibit prominent Br$\gamma$ emission. Stars 13 and 18 also exhibit strong He {\sc i} emission at $\lambda$2.058 {\micron} with a P Cygni profile, and weak He {\sc i} absorption at $\lambda$2.112 {\micron} that may also include a weak N {\sc iii} emission component on the red side of the He {\sc i} absorption line. The Br$\gamma$ line of star 13 exhibits an asymmetry on its blue side owing to a contribution from He {\sc i}. Star 1 exhibits He {\sc i} at $\lambda$2.058 {\micron} in emission, although it is relatively weak. Star 5 exhibits the He {\sc i} $\lambda$2.058 {\micron}  feature as well, but it is in weak absorption.  The spectra of stars 5 and 1 exhibit weak emission from the Na {\sc i} doublet near $\lambda$2.209 {\micron}, and in the case of star 5, weak Fe {\sc ii} emission near $\lambda$2.09 {\micron}. The presence of these low-ionization emission features indicates that
stars 1 and 5 are probably of the B spectral class, with subtypes in the range of B0--B2 (e.g., see Hanson et al. 1996). Their classification as supergiants is justified in Section 4.2. By comparison, stars 13 and 18 must be hotter, as indicated by the presence of He {\sc i} absorption, while the strong P Cygni emission lines of He i at $\lambda$2.058 {\micron} indicate relatively strong winds. Based on these features, we conclude that stars 13 and 18 can be classified as P Cygni-type O supergiants."

\subsection{Strong-lined WN Stars}
Figure 4 presents spectra of stars 11, 17, and 19. These stars exhibit broad emission lines, indicative of fast, extended stellar winds. For star 11 spectral emission features include Br$\gamma$, He {\sc i} at $\lambda$2.058 {\micron} and $\lambda$2.112 {\micron}, and He {\sc ii} at $\lambda$2.189 {\micron}. The latter of these features is probably the superposition of a He {\sc ii} $\lambda$2.189 {\micron} P Cygni profile and a blend of He {\sc i} emission at $\lambda$2.185 {\micron}. Weak emission from N {\sc iii} is also present at  $\lambda$2.247 {\micron}, as well as a weak C~{\sc iii}/N~{\sc iii} blend at $\lambda$2.104 {\micron}. The weakness of He {\sc ii} relative to Br$\gamma$ and He {\sc i} at $\lambda$2.112 {\micron}, and the strong P Cygni emission component of He {\sc i} at $\lambda$2.058 {\micron}, are characteristics of late-type, hydrogen-rich, nitrogen-type stars, specifically WN8--9h stars (\ca{fig97}\cy{fig97}; \ca{mar08}~\cy{mar08}).

The dominant emission feature in the spectrum of stars 17 and 19 are He {\sc ii} at $\lambda$2.189 {\micron}, which indicates that they are early WN (WNE) subtypes. Weak but broad He {\sc i} emission at $\lambda$2.058 {\micron} is present with a deep and highly blueshifted P Cygni absorption component. The large blueshift of this absorption component relative to the peak of the emission component underscores the high wind velocity of star 17 ($\delta\lambda$=0.015 {\micron} $\Rightarrow$ $v_{\textrm{\tiny{wind}}}\approx2200$\,km\,s$^{-1}$) relative to star 11 ($\delta\lambda$=0.006 {\micron} $\Rightarrow$ $v_{\textrm{\tiny{wind}}}\approx900$ km s$^{-1}$). Furthermore, the emission lines of stars 17 and 19 are exceptionally broad, having FWHM$\approx$160 \AA. According to \ca{cro06}~(\cy{cro06}), WN stars with FWHM(He {\sc ii} $\lambda$2.189 {\micron})$~\ge~$130 {\AA} are classified as broad-lined WN stars (WNb). Stars 17 and 19  satisfy the broad criterion, while the relative strength of He {\sc ii} relative to Br$\gamma$ and He {\sc i} is consistent with a WN5 spectral type. Thus, we classify stars 17 and 19 as a WN5b stars. 

\subsection{WC Stars}
Figure 5 presents the $K$-band spectra of star 8, 14, 3, 2, and 4. Each of these stars exhibit emission from C {\sc iii}  at $\lambda$2.115 {\micron}, He {\sc i}, and He {\sc ii} near $\lambda$2.058 {\micron}, C {\sc iii}, and C {\sc iv} near $\lambda$2.07--2.08 {\micron}, and weak He {\sc ii} emission near $\lambda$2.189 {\micron}. These spectral characteristics are consistent with those of late, carbon-type (WC) WRs. A subtype diagnostic is provided by the strength ratio of C {\sc iv} at $\lambda$2.08 {\micron} relative to C {\sc iii} at $\lambda$2.115 {\micron} (\ca{fig97}~\cy{fig97}), which is less than unity for all stars, consistent with WC9 subtype. For reference, we also include our spectrum of the WC9 star WR\,101q (Homeier et al. 2003) in Figure\,5, which we observed on 2008 May 16 with the AAT/IRIS2.

\section{Photometric Properties}
\subsection{Infrared Excess}
Luminous WR and O stars generally exhibit infrared excess from the free-free emission generated in their ionized winds, and also from strong line emission (\ca{had07}~\cy{had07}; \ca{mau09b}~\cy{mau09b}). Thermal continuum excess from hot dust emission is also not uncommon to late-type WC (WCL) stars. Color-color diagrams are useful in assessing the infrared excess of WRs and the likely causes for it. Figure 6 shows a color-color diagram of $H-K_s$ versus $J-H$ for the stars in our sample that were detected in all three near-infrared bands. The interstellar reddening vector is illustrated by the linear trend traced by random field stars that lie toward the GCR, most of which are presumably late-type giants. Stars 1 and 7 lie near the lower left of Figure 6, well separated from most of the other massive stars. The fact that they suffer relatively low extinction implies that they are in the foreground toward the GCR. They also do not appear to exhibit evidence for significant, intrinsic excess. 

While the OB and WN stars in the GCR have color excesses typical for their class, several WC stars in Figure 6 appear to have color excesses that are significantly higher than those of the O and WN stars and 1 WC star (note that only four of the WC stars from our sample are included in Figure 6, since two others have not been detected in the $J$ band). The relatively strong excess emission from WC stars 2 and 4 indicates that there is an additional excess component, which is likely the result of thermal emission from hot dust. Late-type WC stars (WC8--9) commonly exhibit thermal emission from hot dust, and it is generally accepted that the production of dust from these stars is a consequence of binary evolution, whereby the dust is created via the collision of hydrogen-rich and carbon-rich winds from the respective OB and WC stars ({\ca{wil05}~\cy{wil05}}). Several likely examples of this phenomenon are already known in the GCR, and they are typically detectable as hard X-ray sources (e.g., see \ca{mau10a} \cy{mau10a} and references therein). Since stars 2 and 4 were not detected in X-ray surveys of their respective fields (\ca{mun09}~\cy{mun09}),  there is currently no evidence that they are members of colliding-wind binary systems. Nonetheless, the relatively strong excess emission from stars 2 and 4 is consistent with thermal dust emission. Thus, we tentatively classify stars 2 and 4 and WC9?d stars. 

\subsection{Extinction and Luminosity}
We derived the extinction and absolute photometry for each source by comparing the observed $J-K_s$ and $H-K_s$ colors with the intrinsic colors of stars of the same spectral types, which were adopted from the literature (\ca{mar06}~\cy{mar06}; \ca{cro06} ~\cy{cro06}). We used the extinction relation of \ca{nish06}~(\cy{nish06}) for stars near the Galactic center, which is given by \[A_{K_{s}}=1.44\pm0.01E_{H-K_{s}} \] and \[A_{K_{s}}=0.494\pm0.006E_{J-K_{s}}, \] where $E(H-K_{s})=(H-K_{s})_{\textrm{\scriptsize{obs}}}-(H-K_{s})_{0}$, and similarly for the $J-K_s$ colors.  The two values were then averaged to obtain a final extinction estimate, $\overline{A_{K_s}}$. The results are listed in Table 3. The values are similar to other known massive stars in the GCR (e.g., see \ca{mau10a}~\cy{mau10a}). As expected from their positions in Figure 6, stars 7 (O4--6I) and 1 (B0I--B2I) have relatively low $A_{K_s}$ values of 1.16 and 0.95 mag, respectively, which indicates that they lie in the foreground.

Assuming a distance of 8 kpc for the Galactic center (\ca{reid93}~\cy{reid93}), we derived $M_{K_s}$ from the extinction-corrected values. To calculate the bolometric luminosities of the stars, bolometric corrections for the appropriate spectral types were adopted from the same literature sources used to obtain the intrinsic colors, except for the WN8--9h stars, whose bolometric corrections were derived from the $M_K$ and luminosity values of the WN8--9h stars in \ca{mar08}~(\cy{mar08}). These results are also listed in Table 3, and they are more or less consistent with luminosities that are typical for these spectral types. The luminosity values derived for the other O and B stars are also consistent with supergiant luminosity class. We note that the luminosity values obtained for WC9?d stars should be met with caution, since thermal dust emission can have a significant effect on the observed infrared colors, leading to an overestimate of the extinction toward the source. Nonetheless, the luminosity values that we obtain are consistent with those of other WC9 stars.  For the foreground O--4I star 7, we assumed that it shares the same value of $M_K$ as star 6, since they appear to be identical in spectral type. Thus, we derive a distance modulus of 12.79 mag (3.6 kpc) to star 7, which places it in the Norma arm of the Galaxy (\ca{church09}~\cy{church09}). For the B0I--B2I star 1 we have no information on its luminosity class, so we are unable to derive its foreground distance. However, its extinction value is very comparable to star 7; so, we could infer that it also lies at roughly the same distance, also in the Norma arm. If so, then the associated distance modulus of 12.79 mag indicates that star 1 has $M_{K_s}=-6.71$, which implies supergiant luminosity class.

\section{X-ray Detections}
The O4--6I stars 6 and 7 have X-ray counterparts in the master catalog of \ca{mun09}~(\cy{mun09}), which contains 9017 X-ray point sources detected in all combined observations of the GCR with the \textit{Chandra X-ray Observatory}. The X-ray photometric data for these two sources is presented in Table 4. The details of the extraction process are described in \ca{mun09}~(\cy{mun09}). Candidate near-infrared counterparts to the population of GCR X-ray sources were identified in \ca{mau09a}~(\cy{mau09a}); the O4--6I star 6 is source 432 from Table 3 of those authors, and is associated with the \textit{Chandra} source CXOGC J174531.4$-$285716. On the other hand, the foreground O4--6I star 7 that is associated with \textit{Chandra} source CXOGC J174537.9$-$290134 is not present in \ca{mau09a}~(\cy{mau09a}), since it was flagged as having erroneous photometry in the SIRIUS near-infrared catalog that was used for that work; so, for this star we relied on photometry from the Two Micron All Sky Survey (2MASS; Cutri et al. 2003). Table 4 includes X-ray source astrometry, total on-source integration time, photon number counts in the hard (2.0--8.0 keV) and soft bands (0.5--2.0 keV), broadband photon flux, hardness ratios for soft and hard energy bands, and the average energy of all detected photons.

In order to constrain the distance and nature of the X-ray sources, it is useful to examine the hardness ratio, defined as HR=$(h-s)/(h+s)$, where $h$ and $s$ are the fluxes in the hard and soft energy bands, respectively. The soft color, HR0, is defined by letting $h$ and $s$ be the fluxes in the respective 0.5--2.0 keV and 2.0--3.3 keV energy bands. The hard color, HR2, is defined by letting $h$ and $s$ be the fluxes in the 4.7--8.0 keV and 3.3--4.7 keV energy bands, respectively. Galactic center sources, which typically lie behind a hydrogen absorption column of $N_{\textrm{\tiny{H}}}>4\times10^{22}$ cm$^{-2}$, will have most of their soft X-ray photons absorbed by the interstellar medium (ISM), and will, thus, exhibit $\textrm{HR0}\ge -0.175$ (\ca{mau09a}~\cy{mau09a}). Star 6, with HR0=0.56, definitely satisfies the criterion for Galactic center location, while star 7, with HR0=$-$0.17 barely makes the cut.  Compared with HR0, the more energetic photons used to calculate HR2 suffer significantly less absorption from intervening gas and dust, so they are more useful for constraining the intrinsic X-ray hardness. We compared the HR2 colors of these sources with those of the several dozen X-ray emitting massive stars in the GCR that were presented in \ca{mau10a}~(\cy{mau10a}; their Figure 11). Star 6 is among the faintest X-ray sources that has a confirmed infrared counterpart. It was not significantly detected at energies above 4.7 keV, so its HR2 value is $-1.0$. This contrasts with most of several dozen known, massive stellar X-ray sources in the GCR, many of which have significant detections at energies up to 8.0 keV. This feature of the general population is presumed to be the result of emission from a relatively hard thermal plasma with $kT\gtrsim2$ keV. By contrast, the practically null detection of X-rays above 4.7 keV energy from star 6 implies that it is a relatively faint and soft X-ray source ($kT<1$ keV, if it is thermal). Unfortunately, there are not enough counts to effectively model the X-ray spectrum. But if we assume a thermal plasma and guess a reasonable temperature, we can at least get an order-of-magnitude estimate of the X-ray luminosity. To do this, we used the derived extinction values in Table 3 to calculate the corresponding hydrogen column density ($N_{\textrm{\tiny{H}}}$), using the $A_V/A_K\approx16.1$ ratio from \ca{nish08} (\cy{nish08}), assuming a negligible difference between $A_K$ and $A_{K_s}$, and the $A_V/N_{\textrm{\tiny{H}}}$ ratio of \ca{ps95} (\cy{ps95}), which gives $N_{\textrm{\tiny{H}}}/A_{V}=1.8\times10^{21}$ cm$^{-2}$ mag$^{-1}$. We obtained $N_{\textrm{\tiny{H}}}=5.9\times10^{22}$ cm$^{-2}$ for star 6, which is consistent with the canonical average value of $N_{\textrm{\tiny{H}}}=6\times10^{22}$ cm$^{-2}$ that is typical for Galactic center X-ray sources. Using the observed 0.5--8.0 keV energy flux of $F_{X}=7.3\times10^{-16}$ erg s$^{-1}$ cm$^{-2}$ for this source, we employed the \program{PIMMS}{\footnote{http://heasarc.gsfc.nasa.gov/Tools/w3pimms.html}} simulator to calculate the absorption-corrected X-ray flux. We assumed an intrinsic thermal plasma energy value range of $kT=0.6$--1 keV, which is appropriate; after all,  the faintness of this source and its lack of high energy emission (low HR2 value), compared with the other GCR X-ray sources of \ca{mau10a}~(\cy{mau10a}), imply that the plasma temperature is unlikely to be any hotter than this. \program{PIMMS} predicts an unabsorbed X-ray flux in the range of $(1.7$--$9.2)\times10^{-14}$  erg s$^{-1}$ cm$^{-2}$ for star 6. Accounting for the distance of 8 kpc, this implies an X-ray luminosity in the range of $(1.3$--$7.0)\times10^{32}$ erg s$^{-1}$ in the 0.5--8.0 keV energy band. Thus, assuming $kT$ is between 0.6 and 1 keV, the X-ray to bolometric luminosity ratio of star 6 is in the range of $\textrm{log}~(L_X/L_{\textrm{\tiny{bol}}})=-7.2$ to $-6.4$.

The other O4--6I star, 7, has a relatively large X-ray flux in the soft energy range (0.5--2.0 keV) compared with star 6 and the other massive stellar X-ray sources in the GCR. It is also relatively bright overall. This is not surprising, since both the photometry and the extinction value that we calculated for the stellar counterpart in Table 3 imply a foreground distance of 3.6 kpc. So, if the X-ray-detected O4--6I stars 6 and 7 have the same intrinsic nature, then we should expect the foreground star 7 to appear brighter.   Fortunately,  star 7 has enough counts to warrant a fitting of its X-ray spectrum, which can better constrain the physical parameters of the X-ray emission. To do this, we used the program \program{XSPEC} (\ca{ar96}~\cy{ar96}). First, the X-ray spectral data were binned to have at least 30 counts in each energy bin. Then, we performed least squares fits to the data by trying out two different emission models, including a power-law model and a thermal plasma model. The power-law fit required an unrealistically steep photon index of $\Gamma\approx5$, so we do not regard this as a viable model for star 7. However, the thermal plasma model fit the data reasonably well. The parameters fit by the model were $N_{\textrm{\tiny{H}}}$, $kT$, and a normalization factor that is proportional to the emission measure. For the iterative fitting procedure, we used initial guess values of $N_{\textrm{\tiny{H}}}=3.4\times10^{22}$ cm$^{-2}$, derived from the infrared photometry, and $kT=1$ keV. After 30 fitting iterations, \program{XSPEC} returned values of $N_{\textrm{\tiny{H}}}=(3.0\pm0.4)\times10^{22}$ cm$^{-2}$, $kT=0.9\pm0.2$ keV, and an energy flux of $F_X=4.2\times10^{-15}$ erg s$^{-1}$ cm$^{-2}$. The reduced $\chi^2$ of the fit was 0.7 for 13 degrees of freedom. Using these parameters we calculated an absorbtion-corrected X-ray flux of $F_{X}^{\textrm{\tiny{unabs}}}=(5.8\pm1.3)\times10^{-14}$ erg s$^{-1}$ cm$^{-2}$. Using the derived distance of 3.6 kpc, we calculated an X-ray luminosity of $L_{X}=1.3\times10^{32}$ erg s$^{-1}$ for star 7, which implies an X-ray to bolometric luminosity ratio of $\textrm{log}~(L_X/L_{\textrm{\tiny{bol}}})=-7.2$,

The estimated plasma temperatures and X-ray luminosities of stars 6 and 7 are consistent with the values that have been measured for the GCR O4--6I stars in the sample of \ca{mau10a}~(\cy{mau10a}), and the O4--6I stars in the Cygnus OB2 sample of \ca{ac07}~(\cy{ac07}). O stars typically exhibit $\textrm{log}~(L_X/L_{\textrm{\tiny{bol}}})\approx-7$, and this is true for both single stars and those in massive binaries (\ca{osk05}~\cy{osk05}). This relation is a result of the connection between the strength of radiatively driven winds and the temperature of the resulting shocks, which may occur inside the stellar wind via turbulence-induced micro shocks, or from wind collision if the O star is in a massive binary. However, the plasma temperatures reached via the inter-wind micro-shock process typically have $kT<1$ keV, while shocks in colliding-wind binaries generate higher temperatures ($kT\gtrsim1$--2 keV) and produce harder X-ray spectra. Our derived value of $kT=0.9\pm0.2$ keV is insufficient to effectively discriminate between these two mechanisms. Although the average photon energy of both sources is $\langle E \rangle=2.5$--3.0 keV, these photons could simply be the tail of a much softer intrinsic spectrum whose lower energy photons were absorbed by the ISM. Alternatively, our assumption of a single temperature plasma for star 7  might be too simplistic, and the highest energy photons detected from this source could actually be from an intrinsic hard component with $kT\gtrsim1$--2 keV. But without more counts in the softer energy bands, we cannot reliably fit a multi-temperature plasma model to the data, which is generally the case for high X-ray flux O stars in Cygnus.  In conclusion, the X-ray properties of stars 6 and 7 are consistent with thermal emission from either single O stars or those in colliding-wind binaries, although we currently have no good reason to invoke binarity for either source. 

\section{Discussion} 

\subsection{How Complete is the WR Sample in the Central $\approx$90 Parsecs of the Galaxy?}
There are currently 92 identified WRs in the GCR, including all known WNE (WN4--6), WNL (WN7--9), WCE (WC4--7) and WCL (WC8--9) stars identified in the field (including those identified outside of the P$\alpha$ survey area), and those identified in the Arches, Quintuplet, and Central clusters. Their subtype statistics are presented in Table 5. The most noteworthy characteristic of the WR subtype distribution is the high late-type to early-type ratio for both WC and WN stars; out of the 55 WN stars, only 5 are WNE types; and out of 36 WC stars, only 1 is a WCE type. WC stars comprise $\approx$40\% of the total WR population ($\approx$47\% if we exclude the Arches cluster, which is too young to have produced WC stars). 

Figure 7 shows a plot of P$\alpha$ line strength (defined here as the F187N/F190N flux ratio) versus $K_s$-band magnitude for known WRs in the GCR and the remaining unidentified P$\alpha$-excess candidates in the survey area (\ca{wang10}\cy{wang10}; Dong et al. (in preparation). The known sources in Figure 7 include emission-line stars from the GCR field, and from the Arches, Quintuplet, and Central Parsec clusters. The $K_s$-band values of unidentified sources were obtained by cross-correlating the positions of the P$\alpha$-excess candidates with sources from the SIRIUS and 2MASS catalogs (we make a note of caution that a few of the unidentified P$\alpha$ sources having the faintest $K_s$ matches might be spurious associations in this crowded region, and that SIRIUS magnitudes may have significant systematic errors because of source confusion, especially near the Central Parsec Cluster). The data in Figure 7 suggest that all but a few of the remaining unidentified P$\alpha$ candidates occupy a brightness range of $K_s$ and F187N/F190N that is systematically offset from the range occupied by confirmed WRs. Thus, an important question naturally arises as to the completeness of the WR sample at the inner $(l, b)\approx90\times35$ pc$^2$ around Sgr A$^*$ (the survey area). 

WNL stars, which almost completely dominate the current WN sample, have F187N/F190N ratios between 1.2 and 2.9, and have bright counterparts with $K_s\approx7.5$--12 mag. WNL types are expected to generally have $K\lesssim12$ mag when observed in the GCR (e.g., see \ca{fig95}~\cy{fig95}, their Figure 25). Thus, we suspect that only a few WNL stars, at most, could remain unidentified in the survey area.  This also appears to be the case for WNE types; those confirmed thus far, all of which have subtypes in the WN5--6 range, have several of the highest F187N/F190N ratios in the entire sample of WRs. None of the unidentified P$\alpha$-excess candidates have F187N/F190N ratios this large. Earlier WNE stars of subtype WN3--4 have not yet been found in the GCR. Although they are intrinsically the faintest of the WN stars ($M_K\approx-3.1$ mag; \ca{cro06}~\cy{cro06}), having expected $K$-band magnitudes as faint as $\approx$14 in the GCR, WN3--4 stars will exhibit P$\alpha$ excess similar to the WN5--6 stars. Therefore, they should still be easily detectable as relatively strong sources of P$\alpha$ excess\footnote{We have confirmed this statement by examining the $HK$ spectra of WN3--6 and WC4--6 stars obtained from the ground (not shown here). The spectra of the WNE stars exhibit large flux at $\lambda$1.87 {\micron} that can penetrate the severe atmospheric absorption at this wavelength, which implies that such stars will generally have high F187N/F190N ratios.}, modulo the presence of a bright, diffuse background.  Thus, although Figure 7 indicates that several WN stars could remain unidentified, we suspect the WN sample in the survey area is near completion nonetheless. This implies that WNL types heavily dominate the GCR WN population, and that WNE subtypes are intrinsically rare in the region.

The situation is less clear for WC stars. Their F187N/F190N ratios span a range of values similar to the distribution of WNL types, but the faint end of their expected brightness distribution in the GCR extends down to $K\approx15$ mag for the earliest WCE subtypes (\ca{fig95}~\cy{fig95}). This has been confirmed by photometry of the WC5--6 star IRS 3E of the Central Parsec cluster (E58 from \ca{pau06}~\cy{pau06}), which has $K=15.0$ mag and is the only WCE star identified in the GCR to date. Unfortunately, the F187N/F190N ratio for this source could not be measured  because of source confusion and the strong, diffuse background emission from Sgr A West. Even with relatively large intrinsic F187N/F190N ratios, however, sources as faint as IRS 3E will have large photometric uncertainties that could render them weak or undetectable sources of P$\alpha$ excess in the \textit{HST}/NICMOS survey (Dong et al., in preparation). Thus, it is difficult to ascertain the detectability of a hypothetical population of WCE types in the GCR, and we cannot reliably estimate the potential contribution of such stars to the sample of unidentified sources in Figure 7. The same is true for WCL types, which can also be very faint ($K\lesssim14$ mag) and whose line strengths can be heavily diluted in cases where there is significant flux contribution from a host dust continuum. Extremely dusty WCL stars (DWCL; \ca{wil87}~\cy{wil87}), like the Quintuplet proper members, whose emission lines have been completely diluted by a thermal continuum or obscuring dust screen, are not detectable P$\alpha$ excess sources at all; so, some of these stellar types could remain unidentified.  However, DWCL stars are very bright infrared sources ($K_s\approx6$--7 mag) and typically have X-ray counterparts. Based on these properties, the statistics of X-ray/infrared matches from \ca{mau09a}~(\cy{mau09a}), combined with the spectroscopic survey results of \ca{mau10a}~(\cy{mau10a}), imply that most, if not all, DWCL stars in the survey area should have been found already. Still, less extreme WC9d stars with diluted emission lines that are not detectable X-ray sources might remain elusive. 

The near completeness of the WN stars in the survey area implies that apparent dominance of WNL stars over WNE stars is an intrinsic property of the GCR.  Although the sample of WC stars is probably not complete, we also suspect that WCL stars are intrinsically dominant over WCE types. The reasons for this are as follows. It has been well established that WR subtype distributions are correlated with environmental metallicity; this has been established for the Milky Way and other galaxies of the Local Group (\ca{mass96}~\cy{mass96}). WNE and WCE stars appear to be rare in relatively high-metallicity environments ($Z \ge \Zsun$), such as the inner Galaxy, but are common in low-metallicity environments, such as the outer Galaxy ($Z\ge \Zsun$), and the Large Magellanic Cloud ($Z\ll\Zsun$).  This observed trend has been interpreted as a result of the metallicity dependence of WR winds (see \ca{cro07}~\cy{cro07} and references therein for the following discussion). For WN stars, the higher wind density that results from increased metallicity leads to more efficient atomic recombination from high ionization stages (e.g., N$^{5+}$) to lower ions (e.g., N$^{3+}$), which effectively increases the trend toward later spectral types. Alternatively, decreased recombination efficiency and reduced nitrogen content at lower metallicity results in a trend toward earlier WN subtypes. For WC stars in high-metallicity environments, higher wind densities lead to an increased emission strength of  C {\sc iii}  with respect to C {\sc iv}. Since WCL and WCE stars are classified as such based upon the relative strengths of C {\sc iii} and C {\sc iv} emission lines, WC stars in environments with {$Z \ge \Zsun$} will tend to manifest as WCL subtypes (C {\sc iii} stronger than C {\sc iv}). Although this effect was originally suggested by \ca{cro02}~(\cy{cro02}) to be an important factor for the optical classification of WC stars (specifically for C {\sc iii} at $\lambda$5696 \AA), the effect of elevated C {\sc iii} relative to C {\sc iv} for increasing wind density is likely to be mirrored in the infrared spectrum of WC stars as well. Thus, the dominance of WNL and WCL stars over WNE and WCE stars in the GCR is a natural consequence of the fact that the metallicity of the region is at least solar (e.g., see \ca{dav09}~\cy{dav09}, and references therein), if not supersolar (\ca{cah07}~\cy{cah07}; \ca{naj09}~\cy{naj09}).

\subsection{Rogue Massive Stars from the Arches and Quintuplet Clusters}
Figures 8 and 9 show the distribution of ``isolated" massive stars very near the Arches and Quintuplet clusters. For the Arches, there are three WNh stars that lie within $\approx$2 pc, in projection, of the cluster. The magnitudes and spectral types of these outer WNh stars are very similar to the brightest Arches members, some of which have estimated initial masses above 100 {\Msun} (\ca{fig02}~\cy{fig02}). Based on the spectral similarities and the close spatial proximity, we suggest that the three massive stars surrounding the Arches cluster are coeval with the system. The main uncertainty is with regard to their current kinematical state.  One might expect that mass segregation via dynamical friction should concentrate massive stars like these toward the cluster center. However, higher order processes occurring at the cluster center might result in an alternative effect. Indeed, the Arches cluster is the most dense system known in the Galaxy ($3\times10^{5}$ {\Msun} pc$^{-3}$; \ca{fig99b}~\cy{fig99b}). As such, the central region of this cluster must be the site of frequent  gravitational encounters between massive stars and binaries, including binary captures, disruptions, and exchanges. Through such interactions, one might expect a few massive stars to be gravitationally ``kicked" from time to time beyond the tidal radius, which is $\approx$1 pc for the Arches and Quintuplet (\ca{pz02}~\cy{pz02}). This could be the case for  stars 11 and 12 of this work, and G0.10$+$0.02 from \ca{cot99}~(\cy{cot99}), all of which lie well beyond the Arches cluster's tidal radius. Given that the internal velocity dispersion of the Arches cluster is $\sim$10 km s$^{-1}$ (\ca{fig02}~\cy{fig02}), a massive star would not require an unusually large gravitational ``kick" to escape the cluster. Alternatively, one or more of these stars could have formed at the same time and from the same cloud as the main cluster but never became bound to the system. If this is the case then the stars could simply be following a path that is in tandem with that of the cluster, without orbiting the cluster center. These possibilities may also be applicable to the massive stars surrounding the Quintuplet cluster.  \ca{mau07}~(\cy{mau07}) suggested that the O6If$^+$ X-ray source CXOGC J174617.0$-$285131, a suspected colliding-wind binary, may have originated in the Quintuplet, while the same might be true for the nearby LBV G0.120$-$0.048 (\ca{mau10b}~\cy{mau10b}).

Observationally distinguishing between the various possible origins of the stars surrounding the Arches and Quintuplet clusters would be very difficult using radial velocities alone, since both ejected stars \textit{and} comoving stars that were never bound to the clusters would exhibit similar radial velocities, perhaps differing on the order of only 10 km s$^{-1}$. Proper motion measurements, however, would allow for the derivation of a three-dimensional space velocity vector, which would better elucidate the association between these stars and the clusters, and could provide evidence for ejection, specifically. Such measurements are within the capabilities of 10 m class telescopes utilizing adaptive optics. For example, \ca{stolte08}~(\cy{stolte08}) used proper motion measurements over a 4.3 year baseline to derive a three-dimensional space velocity of $232\pm30$ km s$^{-1}$ for the Arches cluster.  As coeval siblings of the Arches cluster stars, we might expect stars 11, 12, and G0.10$+$0.02 to have a similar velocity, whether they were ejected from the cluster or are just comoving with it. Multiple epoch measurements with longer time baselines, however, could discriminate between ejection and comoving scenarios by achieving the required accuracy to obtain a velocity \textit{difference} vector between the individual star vectors and the average cluster vector. If the stars were ejected from the cluster, we would expect these difference vectors to point away from the cluster. 

\subsection{Are there Unidentified Stellar Clusters in the GCR?}
The continuum-subtracted P$\alpha$ mosaic of the inner $(l, b)\approx90\times35$ pc$^2$ of the GCR from \ca{wang10}(\cy{wang10}) is shown in Figure 10, where the positions of the known clusters and the ``isolated" massive stars are illustrated. Aside from the several stars potentially associated with the Arches and Quintuplet clusters, some of the other new discoveries appear to reside in regions that were heretofore unidentified as potential associations. There is a noteworthy association of 3 WC9 stars, a WN7, and an OB supergiant at $(l, b)\approx(359\fdg68--359\fdg76, -0\fdg05)$. Several promising unidentified candidates having $K_s\approx12.4$ and $\approx12.7$ mag and F187N/F190N ratios consistent with them being WC or OB stars also lie in this region. This area is far from any of the known clusters, so, these stars might be the brightest representative members of a bona fide association. There is a more highly concentrated group containing a WC9d, an Ofpe/WN9, and a B2 Ia star at $(l, b)\approx(359\fdg86, 0\fdg00)$, originally identified as a potentially coeval system by \ca{mau07}(\cy{mau07}). In addition, there appears to be a noteworthy collection of 2 WC9 stars and an O supergiant near $(l, b)\approx(0\fdg07, -0\fdg06)$, and a significant concentration of WN stars near $(l, b)\approx(359\fdg98, 0\fdg00)$ and $(l, b)\approx(0\fdg05, 0\fdg02)$, including the central stars of the radio H {\sc ii} regions H1, H2, H5, and H8 (\ca{zhao93}~\cy{zhao93}).

Could these  associations of massive stars represent the ``tip of the iceberg" of stellar clusters that are less compact than the Arches and Quintuplet? There is no measured stellar density enhancement in the vicinity of isolated massive stars in the GCR (\ca{wang10}~\cy{wang10}; Dong et al., in preparation). But because of the strong tidal field in the GCR, stellar clusters like the Arches and Quintuplet will dissolve on a timescale of $\approx$10 Myr (\ca{kim99}~\cy{kim99}), while the time it takes for less extreme stellar clusters to fade into the background may be shorter than this. According to \ca{pz02}~(\cy{pz02}) the cluster dissolution timescale is a sensitive function of galactocentric radius, initial cluster mass, and initial density, such that less compact stellar clusters could become invisible as stellar density enhancements on a shorter timescale than the Arches or Quintuplet ages, or could even persist as undetectable entities for their entire lives. In such cases, an Earth-based observer undertaking a narrowband imaging survey of the GCR might see a small group of evolved emission-line stars with no obvious surrounding stellar cluster. This could be the case for the associations that we have noted above. 

As an alternative explanation for the ``isolated" massive stars, there may be two modes of massive star formation operating in the GCR, one that produces dense starburst systems like the Arches and Quintuplet, and one that produces massive stars in small, sparse associations or in total isolation. Currently, we cannot discriminate between cluster and isolated star formation origins; but continued surveys for coeval massive stars in the vicinity of the new emission-line stars could provide inroads to solving this problem.

\begin{acknowledgements}
The project is partly supported by NASA through a grant HST-GO-11120 from the Space Telescope Science Institute, which is operated by the Association of Universities for Research in Astronomy, Inc., under NASA contract NAS 5-26555.
\end{acknowledgements}

\setlength{\tabcolsep}{0.06in}
\renewcommand{\arraystretch}{1.05}
\begin{landscape}
\begin{deluxetable}{llrrrrrrc}
\tablecolumns{9}
\tablewidth{0pc}
\tabletypesize{\scriptsize}
\tablecaption{Basic Data for New Emission-line Stars}
\tablehead{
\colhead{No.} & \colhead{Name} & \colhead{$\textrm{R.A.}$} & \colhead{$\textrm{Decl.}$} &  \colhead{$J$} & \colhead{$H$} & \colhead{$K_s$} & \colhead{F190N} & \colhead{F187N/F190N} \\ [2pt]
\colhead{} &\colhead{}  & \multicolumn{2}{c}{(deg, J2000)}& \colhead{(mag)} & \colhead{(mag)} & \colhead{(mag)} & \colhead{(\textrm{\scriptsize{mJy}})} & \colhead{}}
\startdata
1  &G359.797$+$0.037    &  266.247738  &   $-$29.090525    & $8.715\pm0.023$ & $7.590\pm0.038$ & $7.030\pm0.013$ &  $1088.79\pm17.10$& $1.28\pm0.03$ \\  
2  &G359.717$-$0.044     & 266.279460&$ -29.200125$&$ 16.842\pm   0.073$&$  13.526\pm   0.019$&$  11.112\pm   0.026$&  $12.58\pm0.20$&$ 1.27\pm0.03$ \\           
3  &G359.746$-$0.090     & 266.341231&$ -29.199841$& \nodata                         &$  14.672\pm   0.029$&$  12.703\pm   0.028$&  $3.75\pm0.06$&$ 1.88\pm0.05$ \\  
4  &G359.691$-$0.072     & 266.290860&$ -29.236897$&$  15.555\pm   0.030$&$  13.019\pm   0.015$&$  11.140\pm   0.018 $& $12.33\pm0.28$ & $1.48\pm0.05$ \\ 
5\tablenotemark{b}  &G359.907$-$0.001\tablenotemark{b}      & 266.350673\tablenotemark{b} &$ -29.016080\tablenotemark{b} $& \nodata                        &$  13.454\pm   0.033\tablenotemark{b}$&$  11.364\pm   0.081\tablenotemark{b}$&  $103.74\pm1.63$&$ 1.17\pm0.03$ \\      
6  &G359.973$-$0.008   & 266.381294&$ -28.954669$&$  15.131\pm   0.021$&$  12.677\pm   0.012$&$  11.366\pm   0.017$&  $15.28\pm0.24$&$ 1.10\pm0.03$ \\      
7  &G359.925$-$0.049   & 266.408263&$-$29.026258 & $11.052\pm0.037$      &$9.578\pm0.044$&$8.898\pm0.119$ &  $185.76\pm2.92$&$ 1.05\pm0.02$ \\  
8  &G359.866$-$0.062      & 266.385493&$ -29.082757$& \nodata                         &$  14.621\pm   0.022$&$  12.016\pm   0.014$&  $5.42\pm0.09$&$ 1.80\pm0.04$ \\          
9  &G0.070$+$0.025         & 266.422032&$ -28.863311$&$  14.799\pm   0.016$&$  11.600\pm   0.013$&$   9.858\pm   0.011$&  $48.43\pm1.08$&$ 1.19\pm0.04$ \\         
10 &G0.058$+$0.014         & 266.426394&$ -28.879828$&$  14.704\pm   0.016$&$  11.666\pm   0.013$&$  10.110\pm   0.028$&  $37.81\pm0.84$&$ 1.33\pm0.04$ \\        
11 &G0.114$+$0.021         & 266.452572&$ -28.828510$& \nodata                          &$  13.612\pm   0.025$&$  11.136\pm   0.013$&  $12.57\pm0.20$&$ 1.90\pm0.04$ \\                          
12 &G0.124$+$0.007         & 266.472510&$ -28.827035$&$  15.367\pm   0.021$&$  12.535\pm   0.017$&$  11.003\pm   0.013$&  $19.20\pm0.30$&$ 1.25\pm0.03$ \\   
13 &G0.059$-$0.068         & 266.506989&$ -28.920983$&$  13.450\pm   0.014$&$  10.687\pm   0.011$&$   9.122\pm   0.046$&  $109.86\pm1.73$&$ 1.22\pm0.03$ \\       
14 &G0.076$-$0.062         & 266.510914&$ -28.903941$&$  16.396\pm   0.038$&$  13.478\pm   0.017$&$  11.608\pm   0.016$&  $8.67\pm0.16$&$ 2.36\pm0.06$ \\        
15 &G0.071$-$0.096         & 266.541810&$ -28.925694$&$  14.984\pm   0.021$&$  12.273\pm   0.017$&$  10.788\pm   0.016$&  $22.50\pm0.35$&$1.27\pm0.03$ \\      
16 &G0.121$-$0.099         & 266.573243&$ -28.884391$&$  14.972\pm   0.018$&$  12.053\pm   0.014$&$  10.459\pm   0.018$&  $29.98\pm0.47$&$ 1.46\pm0.03$ \\      
17 &G0.202$-$0.076         & 266.599325&$ -28.803129$&$  16.330\pm   0.047$&$  13.265\pm   0.011$&$  11.434\pm   0.019$&  $11.68\pm0.18$&$ 2.59\pm0.06$ \\      
18 &G0.238$-$0.071         & 266.615120&$ -28.770077$&$  14.370\pm   0.018$&$  11.291\pm   0.012$&$   9.549\pm   0.042$&  $69.06\pm1.09$&$ 1.28\pm0.03$ \\     
19 & G0.007$-$0.052         & 266.460700&$ -28.957282$&$  15.801\pm   0.031$&$  12.933\pm   0.031$&$  11.339\pm   0.015$&  $13.05\pm0.21$&$ 2.52\pm0.06$ \\          
\nodata & G0.120$-$0.048\tablenotemark{a}         & 266.523436    &    $-$28.858866  & $12.53\pm0.03$ &$9.24\pm0.02$&$7.46\pm0.02$  & 918$\pm$14 & $1.32\pm0.03 $
\enddata
\tablecomments{Positions and $JHK_s$ photometry are from IRSF SIRIUS observations of the Galactic center  (Nishiyama et al. 2006), except for star 1 and 7, which correspond to 2MASS J17445945$-$2905258 and 2MASSJ17453798$-$2901345, respectively (\ca{cut03}~\cy{cut03}).  The \textit{HST/NICMOS} narrowband F187N and F190N measurements are from Dong et al. (in preparation).}
\tablenotetext{a}{This source is the LBV G0.120$-$0.048 from \ca{mau10b}~(\cy{mau10b}). The $JHK_s$ values for this source are from 2MASS. The narrowband measurements reported here are refined values that were extracted from the same data, owing to a recent improvement in the photometry extraction method of Dong et al. (in preparation). In addition, the F190N and F190N/F187N measurements that were presented by \ca{mau10b}~(\cy{mau10b})  for the Pistol Star and qF362 should be updated to $988\pm18$ mJy and $1.21\pm0.03$ for the Pistol Star, and $740\pm13$ mJy and $1.02\pm0.03$ for qF362.}
\tablenotetext{b}{Star \#5 is blended with a neighboring star in IRSF/SIRIUS images; photometry and astrometry unreliable.}
\end{deluxetable}
\end{landscape}

\begin{deluxetable}{lccc}
\tablecolumns{4}
\tablewidth{0pc}
\tabletypesize{\scriptsize}
\tablecaption{Spectroscopic Observations}
\tablehead{
\colhead{Star} & \colhead{Observation Date (UT)} & \colhead{Telescope/Instrument} & \colhead{${\lambda}/\delta{\lambda}$} 
}
\startdata
1   & 2009~Aug~4 07:31   & IRTF/SpeX       & 2000 \\  
2    & 2009~June~13 07:20   & SOAR/OSIRIS   & 1200 \\  
3   & 2009~Aug~4 08:06   & IRTF/SpeX  & 1200 \\ 
4  & 2009~June~13 05:29   & SOAR/OSIRIS    & 1200 \\ 
5   & 2007~Aug~1 08:30   & UKIRT/UIST    & 2133 \\  
6   & 2010~May~24 06:52   & SOAR/OSIRIS      & 1200 \\ 
7  & 2010~May~27 10:23   & SOAR/OSIRIS   & 1200 \\
8   & 2009~Aug~4 07:42   & IRTF/SpeX    & 1200 \\  
9    & 2007~Aug~1 09:05   & UKIRT/UIST    & 2133 \\ 
10    & 2007~Aug~1 06:36   & UKIRT/UIST    & 2133 \\ 
11  & 2008~Jul~31 09:10   & UKIRT/UIST    & 2133 \\ 
12  & 2007~Aug~1 05:57   & UKIRT/UIST    & 2133  \\ 
13   & 2008~Jun~17 09:19   & SOAR/OSIRIS    & 3000 \\ 
14  & 2008~May~16 14:00   & AAT/IRIS2 & 2400 \\ 
15   & 2009~Aug~4 08:40       & IRTF/SpeX        & 1200\\ 
16   & 2008~May~14 19:37   & AAT/IRIS2 & 2400 \\ 
17  & 2008~May~16 15:45   & AAT/IRIS2 & 2400 \\  
18   & 2008~Jun~13 01:10   & SOAR/OSIRIS & 1200 \\ 
19  & 2008~Jul~28 06:48   & UKIRT/UIST & 2133 
\enddata
\end{deluxetable}

\begin{deluxetable}{llrrrcccccc}
\tablecolumns{11}
\tablewidth{0pc}
\tabletypesize{\scriptsize}
\tablecaption{Extinction and Absolute Photometry for Selected Stars}
\tablehead{ \colhead{Star} & \colhead{Spectral} & \colhead{$K_s$} & \colhead{${(J-K_s)}_0$} & \colhead{${(H-K_s)}_0$} & \colhead{$A^{J-K_{s}}_{K_{s}}$} & \colhead{$A^{H-K_{s}}_{K_{s}}$} & \colhead{$\overline{A_{K_{s}}}$} & \colhead{$M_{K_s}$} & \colhead{BC$_K$\tablenotemark{a}} & \colhead{$L_{\textrm{\tiny{bol}}}$} \\ [2pt]
\colhead{} &\colhead{Type} & \colhead{(mag)} & \colhead{(mag)} &\colhead{(mag)}  &\colhead{(mag)}  &\colhead{(mag)}  &\colhead{(mag)}  &\colhead{(mag)} & \colhead{(mag)} &\colhead{(log \Lsun)} }
\startdata
1       & B0I--B2I &   7.03   &                  $-$0.21 &     $-$0.10 & 0.94 & 0.96  & 0.95\tablenotemark{b} &$-$6.71\tablenotemark{b}  & $-$3.7  & 6.06\tablenotemark{b} \\
2         &  WC9?d   &  11.11 &      0.23 &      0.26  &     2.91 &     2.86 &      2.89                   &    $-$6.28    & $-$3.6 & 5.85\tablenotemark{b} \\ 
3        & WC9 &  12.70 &      0.23 &      0.26 &    \nodata &      4.06 &      4.06                   &    $-$5.86&  $-$3.6& 5.68\\ 
4           &WC9?d &11.14 & 0.23 &           0.26  & 2.21 & 2.15  & 2.18                                     & $-$5.54  & $-$3.6 & 5.56\tablenotemark{c}  \\ 
5       &  B0I--B2I\tablenotemark{d}   &11.36\tablenotemark{d}  &     $-$0.21 &     $-$0.10 &  \nodata &     3.16\tablenotemark{d}  &      3.16\tablenotemark{d}                &     $-$6.30\tablenotemark{d} &  $-$3.7\tablenotemark{d} & 5.90\tablenotemark{d}  \\ 
6        & O4--6I  & 11.37  & $-$0.21  &   $-$0.10            & 1.96 & 2.05   & 2.01             & $-5.14$& $-$4.3& 5.68  \\ %
7     & O4--6I   & 8.90  & $-$0.21  & $-$0.10 & 1.17  & 1.14 & 1.16\tablenotemark{b} &  $-5.14$\tablenotemark{b} & $-$4.3&  5.68\tablenotemark{b} \\ %
8       & WC9 & 12.02  & 0.23  & 0.26  & \nodata &  3.12 & 3.12                                        &  $-5.60 $& $-$3.6 &  5.58 \\ %
9        &  O4--6If$^+$   & 9.86  &    $-$0.21 &     $-$0.10   &      2.54 &      2.67 &      2.61   &  $-$7.25& $-$4.3 & 6.52 \\ 
10        & O4--6If$^+$      &10.11 &    $-$0.21 &     $-$0.10 &     2.37 &      2.40 &      2.39    &     $-$6.78& $-$4.3 & 6.33 \\ 
11      &WN8--9h  &  11.14 & 0.13 &      0.11 &  \nodata &      3.41 &      3.41                                &     $-$6.77&$-$4.2 & 6.29\\ 
12  & WN8--9h      &  11.00 &     0.13 &     0.11 &    2.09    &  2.05 &      2.07        &    $-$5.57 & $-$4.2  & 5.81 \\ 
13       & P Cyg-type OI &   9.12 &       $-$0.21 &     $-$0.10 &  2.24 &      2.41 &      2.33                     &     $-$7.71&$-$3.7& 6.46\\ 
 14   & WC9  &  11.61 &   0.23 &      0.26  &    2.41 &      2.14 &      2.28                                  &     $-$5.17&$-$3.6 & 5.41\\ 
 15     & WN8--9h  & 10.79  & 0.13  & 0.11  & 2.01  & 1.98 & 2.00 & $-$5.71 &  $-$4.2 &  5.86 \\ 
16     & WN8--9h     &  10.46 &     0.13 &     0.11 &    2.17 &      2.24 &      2.15 &     $-$6.19&$-$4.2  & 6.06\\ 
17       & WN5b &  11.43 &   0.36 &      0.26 &        2.40 &      2.09 &      2.24                           &     $-$5.31&$-$4.4& 5.54 \\ 
18       & P Cyg-type OI &   9.55      &                  $-$0.21 &     $-$0.10 & 3.37 & 3.35 & 3.36 & $-$8.31 & $-$3.7  & 6.70 \\
19      & WN5b &  11.34 &  0.36 &      0.26  &  2.03 &      1.92 &      1.98                        &     $-$5.31& $-$4.4& 4.96 
\enddata
\tablenotetext{a}{Bolometric corrections were derived from Martins et al. (2008) for WN8--9h stars; from Crowther et al. (2006) for other WRs; and from Martins et al. (2006) for the OB stars.}
\tablenotetext{b}{The relatively low value of extinction for the O4--6I star 7 and B0I--B2I star 1, compared to the other stars in this table, implies that they are located in the foreground. We estimated the distance to 7 (O4--6I) by adopting $M_{K_s}$ and $L_{\textrm{\tiny{bol}}}$ from the similar O4--6I star 6. The resulting distance modulus is 12.79 mag (3.6 kpc), which implies that 7 lies in the Norma arm of the Galaxy (Churchwell et al. 2009). Since 1 has an extinction that is comparable to 7, one might also assume that these stars have comparable distances. If so, this implies  1 has $M_{K_s}=-6.71$ mag, which means it is a supergiant.}
\tablenotetext{c}{These values may be erroneously high. The color-color diagram in Figure 6 implies that these stars may emit thermal excess from hot dust, in which case the extinction cannot be reliably derived using the near-infrared colors.}
\tablenotetext{d}{As noted in Table 1, Star \#5 is blended with a neighboring star in the IRSF/SIRIUS images; photometry unreliable.}
\end{deluxetable}

\begin{deluxetable}{lrr}
\tablecolumns{3}
\setlength{\tabcolsep}{0.1in}
\renewcommand{\arraystretch}{1.4}
\tablewidth{0pc}
\tablecaption{\textit{Chandra} X-ray  Data for G359.973$-$0.008 and  G359.925$-$0.049}
\tablehead{\colhead{Value} & \colhead{G359.973$-$0.008 (Star 6)} & \colhead{G359.925$-$0.049 (Star 7)}}
\startdata
Spectral type  & O4--6I& O4--6I \\
CXOGC J & 174531.4$-$285716 & 174537.9$-$290134\\
R.~A. (J2000) & 266.38108 &266.40829  \\
Decl. (J2000) & $-$28.95466 &$-$29.02625 \\
$\sigma_{\textrm{\tiny{X}}}$ (arcsec.) & 0.5 & 0.3 \\
$C_{\textrm{\tiny{net}}}^{\textrm{\tiny{~soft}}}$ (0.5--2.0 keV) & $10.8^{+9.5}_{-7.3}$ & $140.6^{19.9}_{19.2}$\\
$C_{\textrm{\tiny{net}}}^{\textrm{\tiny{~hard}}}$ (2.0--8.0 keV) &$58.8^{+17.3}_{-23.0}$ & $150.8^{+21.4}_{-21.2}$ \\
$F_{\textrm{\tiny{tot}}}$ (s$^{-1}$ cm$^{-2}$) & $2.44\times10^{-7}$ & $1.37\times10^{-6}$ \\
HR0 & $0.57^{+0.28}_{-0.31}$&  $-0.17^{+0.11}_{-0.10}$ \\
HR2 & $-1.00^{}_{}$ & $-0.07^{+0.26}_{-0.26}$ \\
$\langle{E}\rangle$/photon (keV) & 3.0 & 2.5 \\
$kT$ (keV) & 0.6--1.0 & $0.9\pm$0.2 \\
$N_{\textrm{\tiny{H}}}$ ($10^{22}$ cm$^{-2}$) & 5.9 & $3.0\pm0.4$ \\
Norm. factor & \nodata & $3.2\pm2.3\times10^{-5}$ \\
$F_X^{\textrm{\tiny{unabs}}}$  (erg s$^{-1}$ cm$^{-2}$) &  $(1.7$--$9.2)\times10^{-14}$  & $5.8\pm1.3\times10^{-14}$  \\
$log~L_{X}$ (erg s$^{-1}$) & $(1.3$--$7.0)\times10^{32}$ & $1.3\times10^{32}$ \\
$log~(L_{X}/L_{\textrm{\tiny{bol}}})$ & $-7.2$ to $-6.4$ & $-7.2$ 
\enddata
\tablecomments{The following X-ray data were taken from the catalog of \ca{mun09}~(\cy{mun09}): X-ray counts in the soft ($C_{\textrm{\tiny{net}}}^{\textrm{\tiny{~soft}}}$) and hard ($C_{\textrm{\tiny{net}}}^{\textrm{\tiny{~hard}}}$) energy bands, hardness ratios HR0 and HR2, photon flux $F_{\textrm{\tiny{tot}}}$, and average photon energy $\langle{E}\rangle$. All other values were derived here. For G359.973$-$0.008 (Star 6), the values of $kT$, $N_{\tiny{\textrm{H}}}$, and $F_X^{\textrm{\tiny{unabs}}}$ were derived from the infrared and X-ray photometry. For G359.925$-$0.049 (Star 7), these same values were derived via a thermal plasma model fit, using the program \program{XSPEC} (\ca{ar96}~\cy{ar96}), which also yielded the model normalization factor. Uncertainties are present when available.}
\end{deluxetable}

\begin{deluxetable}{lccccccccc}
\tablecolumns{9}
\tablenum{5}
\tablewidth{0pc}
\tabletypesize{\scriptsize}
\tablecaption{Distribution of WR Subtypes in the GCR}
\tablehead{ \colhead{Group} & \colhead{WNE} & \colhead{WNL} & \colhead{WCE} & \colhead{WCL} & \colhead{WR$_{\textrm{tot}}$} & \colhead{WC/WN} & \colhead{WCL/WC} & \colhead{WC/WR$_{\textrm{tot}}$} & \colhead{References} }
\startdata
GCR ``Isolated"                 & 4    & 15 &    0  & 9 &  28 &  0.47 & 1.00 &  0.32 & 1, 2, 3, 4, 5, 6, 12 \\   
Arches                                 & 0   & 13  &  0 &  0   &  13 & n/a &  0.00   &   0.00 &  7, 11 \\
Quintuplet                           & 1   & 5  &   0  &  14 &  20 & 2.33 & 1.00 &  0.70 & 2, 4, 8, 9  \\
Central                                &  1  & 17 &  1  & 12  &  31 & 0.72 & 0.92  & 0.42 &  10 \\
\textbf{GCR Total}              & 6   & 50  &  1  & 35 &  92 & 0.66  & 0.97 & 0.39 & \nodata   
\enddata        
\tablecomments{References: (1) This work; (2)  Mauerhan et al. (2010a); (3) Cotera et al. (1999); (4) Homeier et al. (2003); (5)  Muno et al. (2006); (6)  Mikles et al. (2006); (7) Figer et al. (2002); (8) Liermann et al. (2009); (9)  Figer et al. (1999); (10)  Paumard et al. (2006); (11) Martins et al. (2008); (12) Hyodo et al. (2008). Note that Stars 11, 12, and G0.10$+$0.02, shown in Figure 8 of the original manuscript, have been designated as ``isolated" stars for this table, even though they are likely to be be associated with the Arches cluster.}
\end{deluxetable}

\newpage

\begin{figure*}[t]
\centering
\epsscale{1}
\plotone{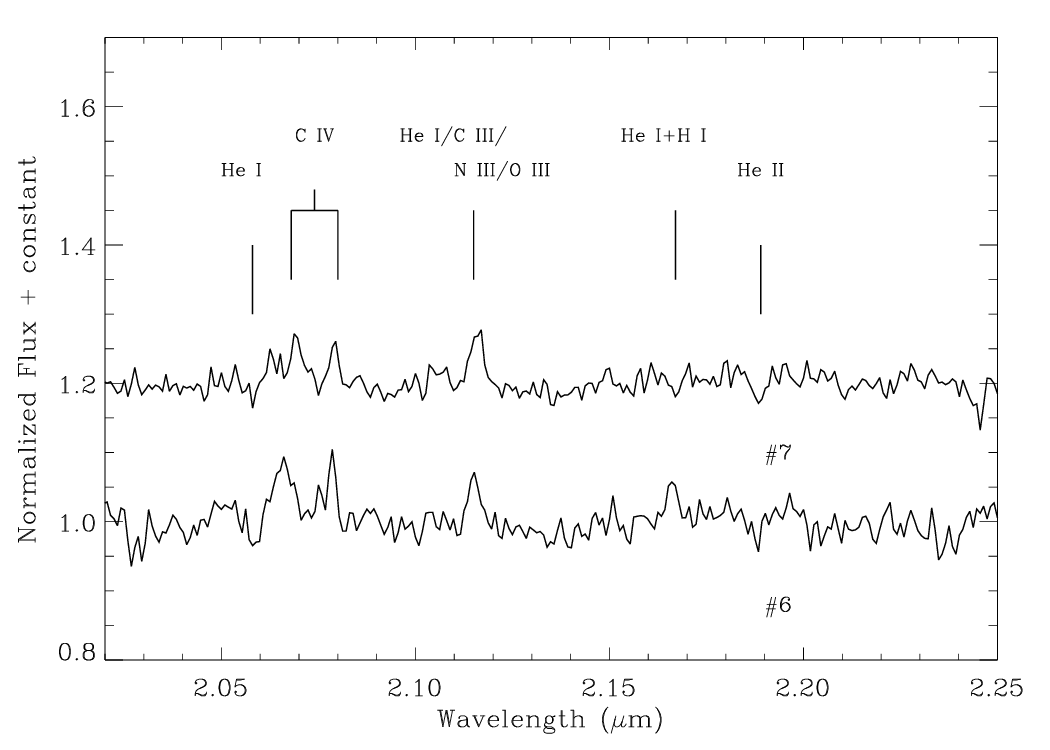}
\caption[]{\linespread{1}\normalsize{$K$-band spectra of confirmed O4--6I stars G359.973$-$0.008 (Star 6) and G359.925$-$0.049 (Star 7), which are counterparts to \textit{Chandra} X-ray sources CXOGC J174531.4$-$285716 and CXOGC J174537.9$-$290134, respectively.}}
\end{figure*}

\begin{figure*}[t]
\centering
\epsscale{1}
\plotone{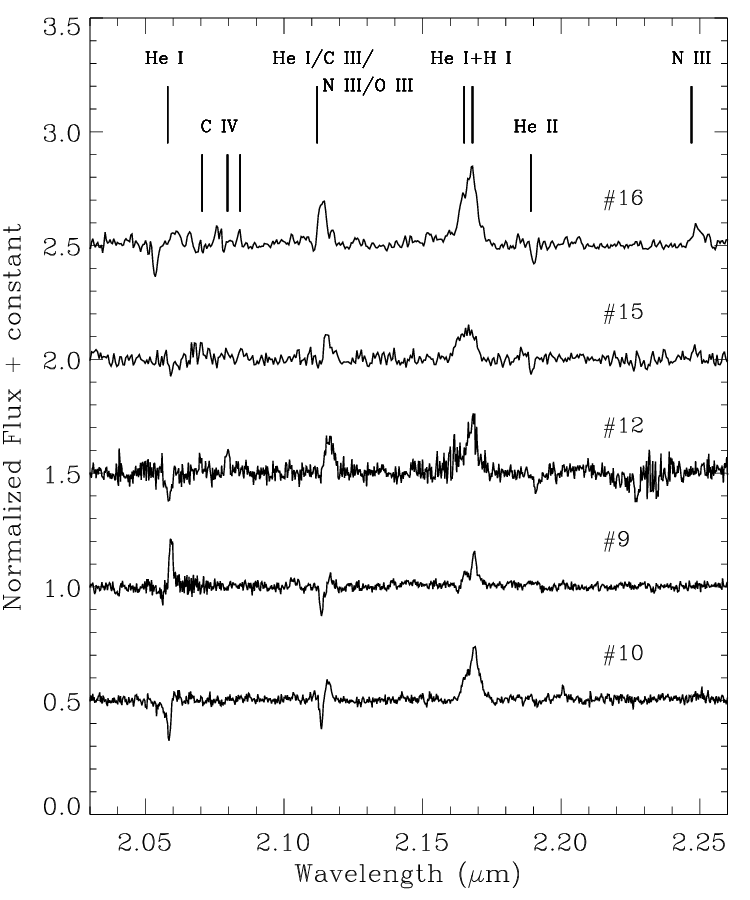}
\caption[]{\linespread{1}\normalsize{$K$-band spectra of confirmed WNh and O4--6If$^{+}$ stars.}}
\end{figure*}

\begin{figure*}[t]
\centering
\epsscale{1}
\plotone{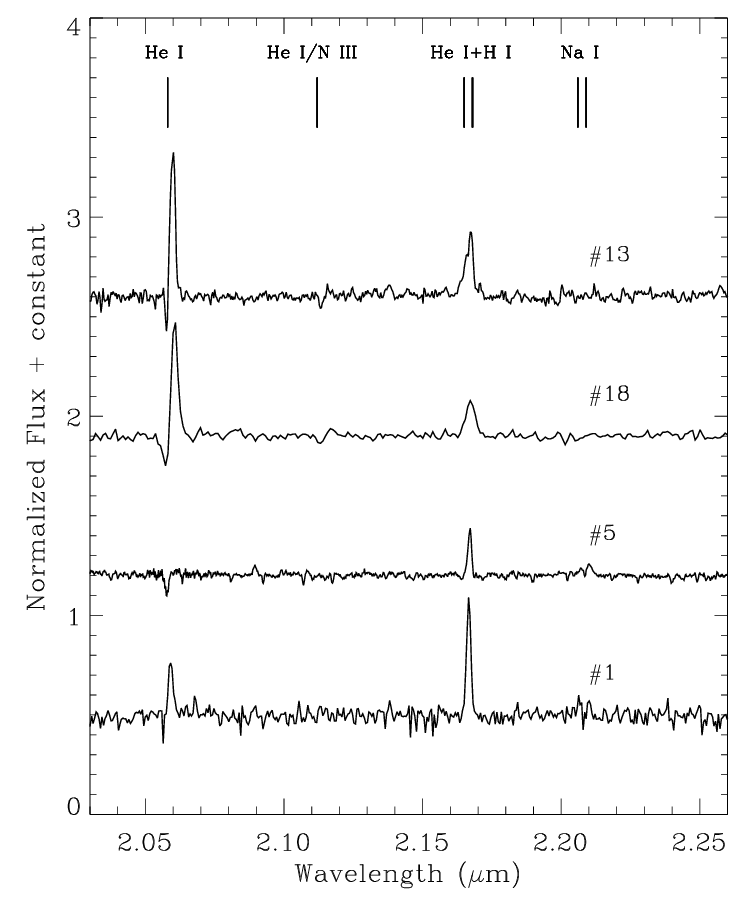}
\caption[]{\linespread{1}\normalsize{$K$-band spectra of P Cyg-type O supergiants and B supergiants.}}
\end{figure*}

\begin{figure*}[t]
\centering
\epsscale{1}
\plotone{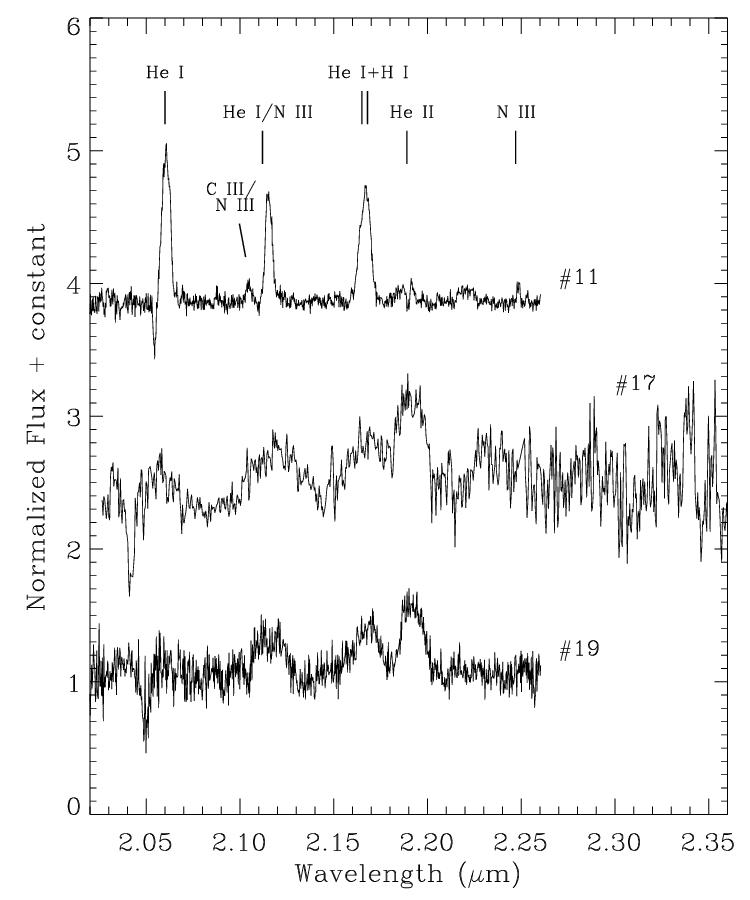}
\caption[]{\linespread{1}\normalsize{$K$-band spectra of confirmed strong-lined WN stars.}}
\end{figure*}

\begin{figure*}[t]
\centering
\epsscale{1}
\plotone{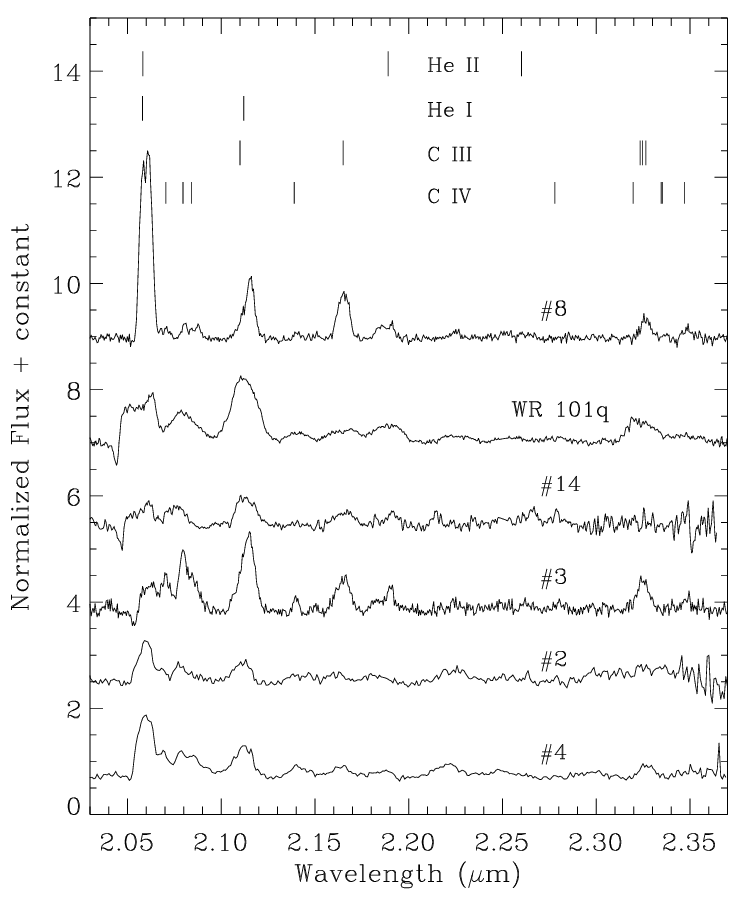}
\caption[]{\linespread{1}\normalsize{$K$-band spectra of confirmed WC9 stars and WR\,101q.}}
\end{figure*}

\begin{figure*}[t]
\centering
\epsscale{1}
\plotone{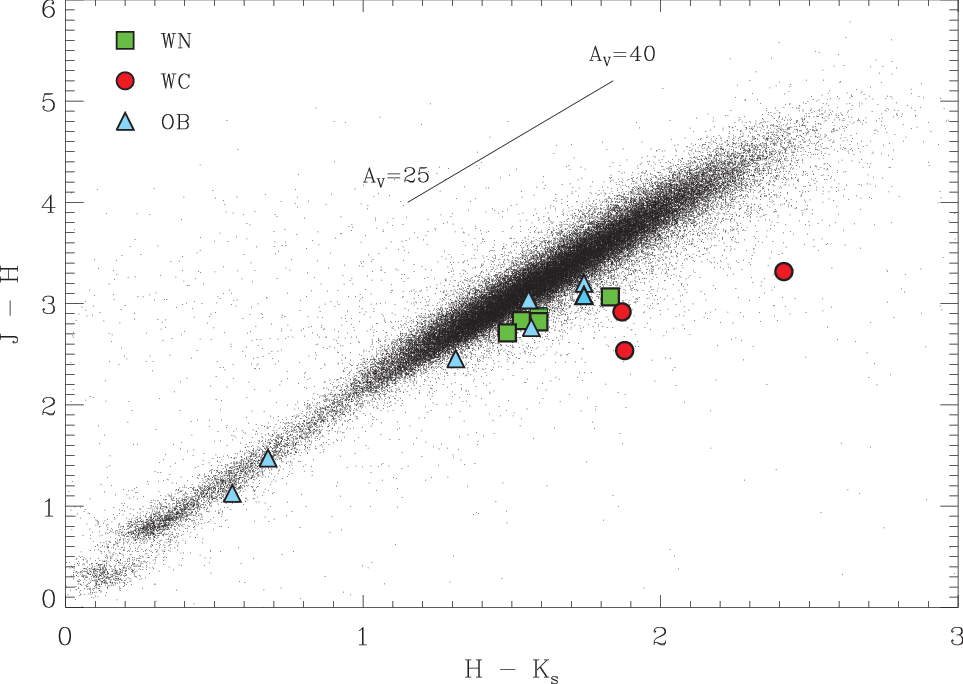} 
\caption[]{\linespread{1}\normalsize{Near-infrared color-color diagram of new WN, WC, and OB stars (green squares, red circles, and blue triangles, respectively}). The small dots are random field stars. All of the massive stars exhibit infrared excess, mainly attributable to free-free and line emission from their dense, ionized winds.  The  WC9 stars may be sources of an additional excess component of thermal emission from hot dust. A color version of this figure is available in the online journal.}
\end{figure*}

\begin{figure*}[t]
\centering
\figurenum{7}
\epsscale{1}
\plotone{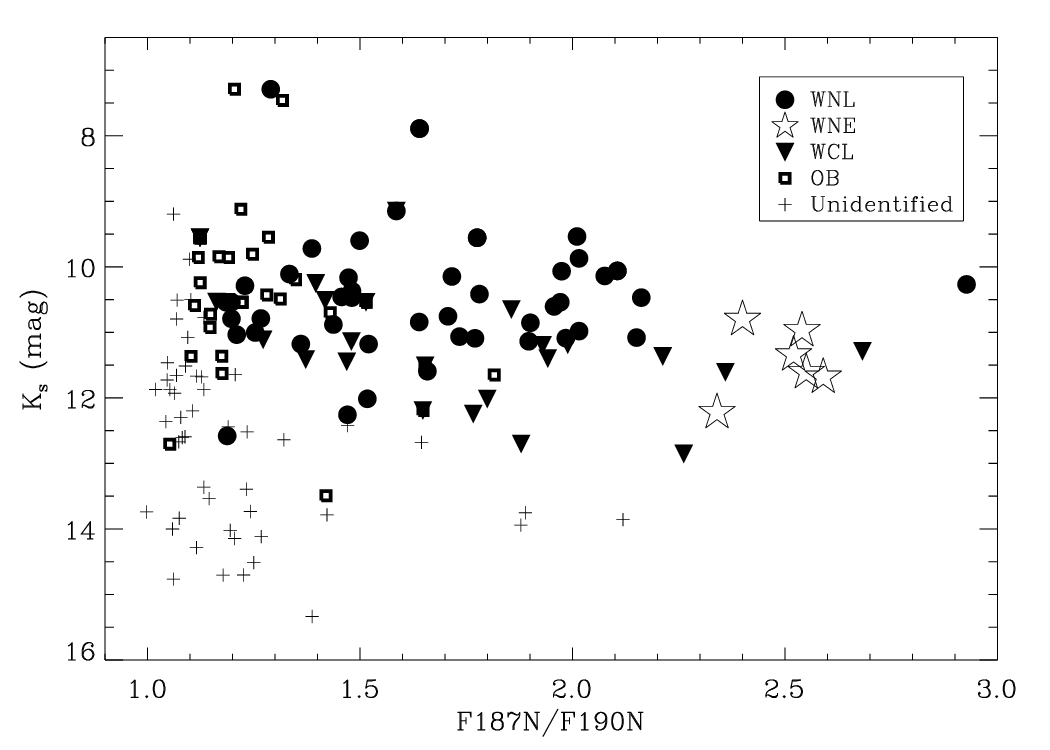}
\caption[]{\linespread{1}\normalsize{Observed line strength (F187N/F190N) versus $K_s$-band SIRIUS magnitude for unidentified candidate P$\alpha$-excess sources (crosses), and for confirmed massive stars. The brightest known sources ($K_s\lesssim8$ mag), which are saturated in the SIRIUS catalog, were plotted using their 2MASS photometry. The figure includes objects from the Arches, Quintuplet, and Central Parsec clusters, and from the GCR field. Stars are marked according to their approximate spectral type, defined in the legend at the upper right of the figure. The brightness distribution of unidentified sources suggests that the sample of WN types in the survey might be near completion.}}
\end{figure*}

\begin{figure*}[t]
\centering
\epsscale{.7}
\plotone{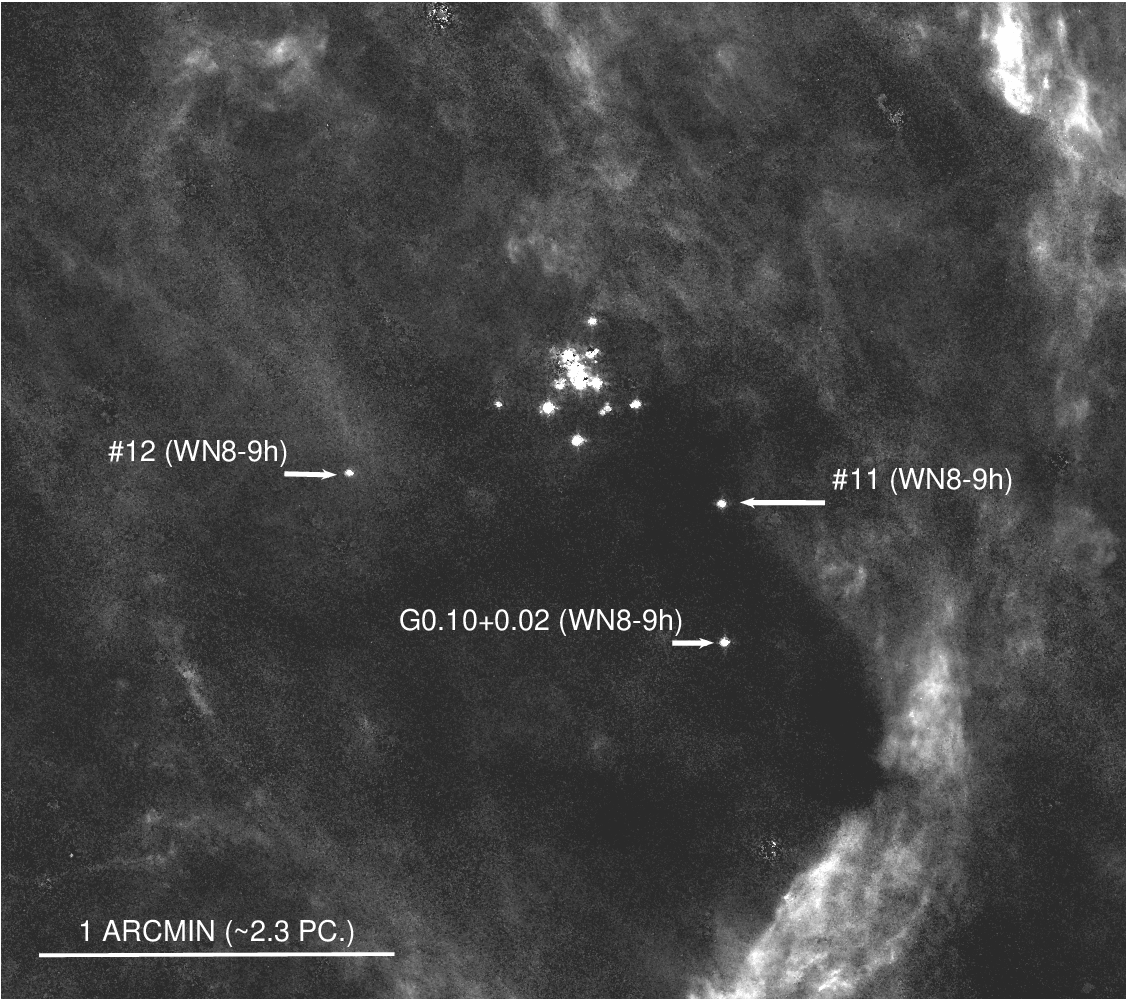}
\caption[]{\linespread{1}\normalsize{\textit{HST}/NICMOS P$\alpha$ survey image from \ca{wang10}(\cy{wang10}) of the Arches cluster region and nearby WNh stars. The marked stars  11 and  12 are new discoveries, while G0.10$+$0.02 was previously reported in \ca{cot99}(\cy{cot99}). The surrounding stars all lie within 1--2 pc, in projection, of the cluster (beyond the tidal radius of $\approx1$ pc), and they may have escaped from it. North is up and east is to the left in this image.}}
\end{figure*}

\begin{figure*}[t]
\centering
\epsscale{.7}
\plotone{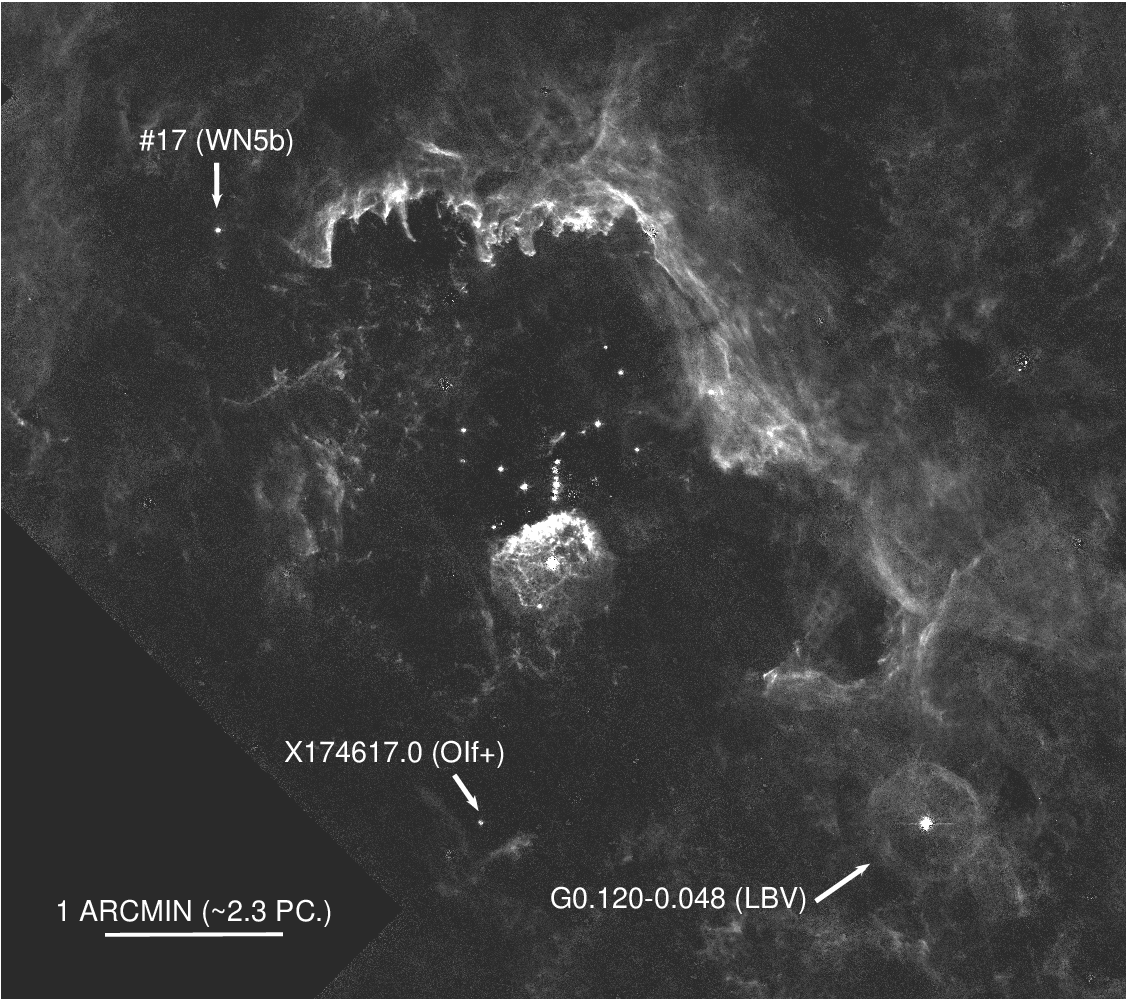}
\caption[]{\linespread{1}\normalsize{\textit{HST}/NICMOS P$\alpha$ survey image of the Sickle region from \ca{wang10}(\cy{wang10}), which contains the Quintuplet cluster and nearby isolated massive stars, including the O6If$^{+}$ X-ray source X174517.0 from \ca{mau07} (\cy{mau07}), the LBV G0.120$-$0.048 from \ca{mau10a}~(\cy{mau10a}), and the new WN5b star  17 of this work. All of the unmarked point sources are known massive stars spectroscopically identified in \ca{fig99}~(\cy{fig99}). North is up and east is to the left in this image.}}
\end{figure*}

\begin{landscape}
\begin{figure*}[t]
\centering
\figurenum{10}
\epsscale{1}
\plotone{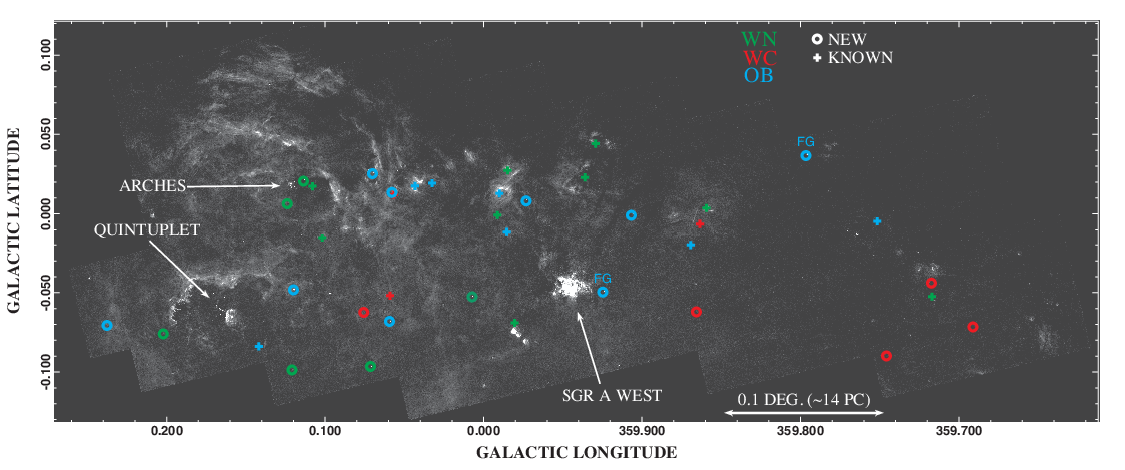}
\caption[]{\linespread{1}\normalsize{\textit{HST}/NICMOS P$\alpha$ survey image of the Galactic center region from Wang et al. (2010). The positions of the isolated massive stars are marked (i.e., those located outside of the known extent of the Arches, Quintuplet, or Central clusters) and color coded according to spectral type (see the image legend). The discoveries of this work (circles) are marked separately from the known sources (crosses). The two foreground O stars are labeled ``FG". Bona fide cluster members are not marked. }}
\end{figure*}
\end{landscape}

\end{document}